\documentclass[conference]{IEEEtran}
\IEEEoverridecommandlockouts
\usepackage[letterpaper, margin=0.75in, top=0.75in, bottom=1in]{geometry}
\usepackage{cite}
\usepackage{amsmath,amssymb,amsfonts}
\usepackage{algorithmic}
\usepackage{graphicx}
\usepackage{textcomp}
\usepackage{xcolor}
\usepackage{subcaption}
\usepackage[font=small,labelfont=bf]{caption}
\usepackage{titlesec}

\newcommand{\EE}{\mathbf{E}}
\newcommand{\HH}{\mathbf{H}}
\newcommand{\XX}{\mathbf{x}}
\newcommand{\NN}{\mathbf{n}}
\newcommand{\YY}{\mathbf{y}}
\newcommand{\FF}{\mathbf{F}}
\newcommand{\SSS}{\mathbf{s}}

\titlespacing*{\section}{0pt}{1ex}{1ex}  
\titlespacing*{\subsection}{0pt}{0.5ex}{0.5ex}  

\def\BibTeX{{\rm B\kern-.05em{\sc i\kern-.025em b}\kern-.08em
    T\kern-.1667em\lower.7ex\hbox{E}\kern-.125emX}}
\begin{document}

\title{Impact of Synchronization Offsets and CSI Feedback Delay in Distributed MIMO Systems\\
}

\author{\IEEEauthorblockN{
         Kumar Sai Bondada\IEEEauthorrefmark{1},
         Daniel J. Jakubisin\IEEEauthorrefmark{1}\IEEEauthorrefmark{2},
         R. Michael Buehrer\IEEEauthorrefmark{1}\\
} 
     \IEEEauthorblockA{
         \IEEEauthorrefmark{1}Wireless@VT, Bradley Department of ECE, Virginia Tech, Blacksburg, VA, USA \\
         \IEEEauthorrefmark{2}Virginia Tech National Security Institute, Blacksburg, VA, USA  
         } 
         }
         

\maketitle

\vspace{-2em}  
\begin{abstract}
The main challenges of distributed MIMO systems lie in achieving highly accurate synchronization and ensuring the availability of accurate channel state information (CSI) at distributed nodes. This paper analytically examines the effects of synchronization offsets and CSI feedback delays on system capacity, providing insights into how these affect the coherent joint transmission gain. The capacity expressions are first derived under ideal conditions, and the effects of synchronization offsets and feedback delays are subsequently incorporated. This analysis can be applied to any distributed MIMO architecture. A comprehensive study, including system models and simulations evaluating the analytical expressions, is presented to quantify the capacity degradation caused by these factors. This study provides valuable insights into the design and performance of distributed MIMO systems. The analysis shows that time and frequency offsets, along with CSI feedback delay, cause inter-layer interference. Additionally, time offsets result in inter-symbol interference.

\end{abstract}


\section{Introduction \& Related Works}
Sixth-generation (6G) wireless networks are envisioned to deliver immersive, hyper-reliable, and low-latency communication, alongside massive and ubiquitous connectivity to meet the demands of emerging applications. Compared to fifth-generation (5G) networks, 6G is anticipated to achieve a 5- to 10-fold improvement in spectral efficiency and peak data rates reaching 1 Tbps \cite{ITU2024, You2021, 9040264}. This significant enhancement is motivated by novel use cases, such as autonomous vehicles, intelligent transportation systems, industrial automation, environmental sensing, and augmented/virtual reality (AR/VR). These advancements necessitate a paradigm shift in communication networks toward distributed and collaborative architectures capable of supporting the stringent performance requirements of 6G.

Distributed MIMO systems are emerging as one of the key enablers for 6G communications \cite{10188355, 10757642}. Prior research has demonstrated the potential of these systems to achieve significant improvements in both spectral efficiency and energy consumption \cite{6601776}. Various distributed MIMO architectures \cite{6495775, 6574665, 6804225} have been proposed under different terminologies, each focusing on objectives such as improving spectral and energy efficiency \cite{5490981}, mitigating inter-cell interference \cite{1678166,4114255, 5594708, 5706317}, ensuring uniform quality of service across the network \cite{7227028,7827017, 10278803}, extending communication range \cite{6735697}, realizing the theoretical gains of massive MIMO by addressing form factor constraints, and offering macro-diversity. Theoretical bounds on capacity by scaling up the number of distributed nodes have been studied in \cite{6415392}. The evolution of distributed MIMO systems has transitioned from connecting distributed mobile radios to a central processing unit via high-speed wired connections to exploring wireless connections enabled by the large bandwidths available in millimeter-wave frequencies \cite{9786576,9048836,10066319}.


Although distributed MIMO systems promise significant improvements in spectral and energy efficiency, they face several challenges, such as achieving highly accurate synchronization and ensuring the availability of accurate channel state information (CSI) at distributed radios (nodes) for coherent joint transmission \cite{9492307,6574665,6804225}. Most prior works assume perfect synchronization or perfect CSI availability at the nodes when analyzing the proposed distributed MIMO systems. However, only a limited number of studies \cite{1417557, 4533899,6250567} have explored the impact of synchronization errors, often considering only timing errors and evaluating their effects through bit error rate (BER) analysis. The work in \cite{4202181} examined the effects of initial phase offset errors in a multiple-transmitter scenario with a single receiver, where each transmitter and the receiver are equipped with a single antenna. While these studies provide valuable insights, they primarily address limited scenarios or focus on specific types of errors, leaving a gap in understanding how synchronization errors and CSI feedback delays jointly impact the capacity of distributed MIMO systems, especially in multi-antenna configurations. The current paper studies the capacity performances of coherent joint transmission by analyzing the impact of synchronization errors and CSI feedback delays on the capacity of distributed MIMO systems with multiple antenna transceivers. Specifically, it quantifies the degradation in capacity caused by these factors and proposes a detailed analysis framework to better understand their implications for distributed MIMO performance. The proposed analysis is applicable to any distributed MIMO architecture.

The remainder of this paper is organized as follows: In Section II, we describe the system model under ideal conditions and analyze the impact of synchronization errors and CSI feedback delay. Section III presents the capacity expressions considering these effects. Section IV provides simulations numerically evaluating the analytical expressions and insights derived from the analysis. Finally, Section V concludes the paper with key findings and discusses potential future research directions.



\textbf{Notation:} Boldface uppercase letters denote matrices; boldface lowercase letters denote vectors.

\section{System Model and Architecture}
\begin{figure}[htpb]
    \centering
    \includegraphics[width=0.8\columnwidth]{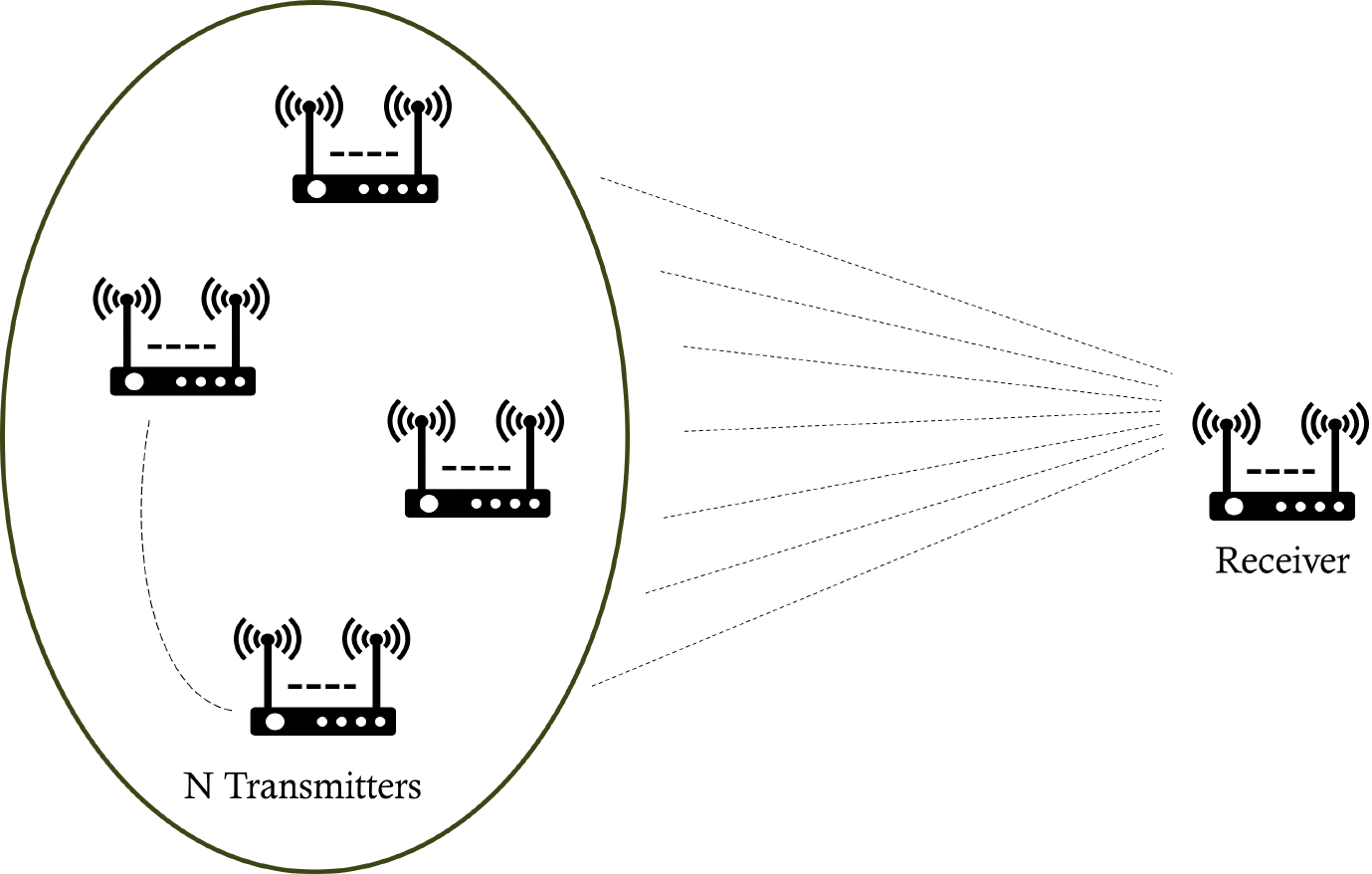}
    \caption{A group of $N$ cooperating transmitters coherently joint transmitting to the receiver.}
    \label{fig:cap_sys_arch}
\end{figure}

We consider a system consisting of multiple $N$ transmitters cooperating in coherent joint transmission to the receiver as shown in Fig. \ref{fig:cap_sys_arch}. All the transmitters have the same transmit power level, and each transmitter and receiver has the same number of antennas. We assume an ideal scenario, where the transmit data and precoder information are available at all the transmitters to focus on the impact of synchronization offsets and CSI feedback delays. For an analysis of capacity accounting for wireless information sharing among the transmitters, see our prior work \cite{10757642}. The received signal at time $t$ can be represented as
\begin{equation}
\YY(t) = \sum_{i=1}^{N} \HH_i(t) \XX_i(t) + \NN(t),    
\label{eq:1}
\end{equation}
where \(\HH_i(t)\) represents the channel matrix between the $i$-th transmitter and the receiver whose elements follow uncorrelated complex Gaussian distribution with zero mean and unit variance, \(\XX_i(t) = \FF_i(t) \SSS(t)\) is the $i$-th transmit signal vector with \(\FF_i(t)\) being the individual precoder of the $i$-th transmitter, \(\SSS(t)\) is the common transmitted data signal vector and \(\NN(t)\) is the additive white Gaussian noise (AWGN) signal vector ($\NN(t) \sim \mathcal{CN}(0, \mathbf{I}_{N_r})$). Substituting \(\XX_i(t)\) into the Eq. \ref{eq:1}, we have
\[
\YY(t) = \sum_{i=1}^{N} \HH_i(t) \FF_i(t) \SSS(t) + \NN(t).
\]
This can also be represented in matrix form as
\begin{multline*}
 \YY(t) = 
\begin{bmatrix}
\HH_1(t) & \HH_2(t) & \cdots & \HH_N(t)
\end{bmatrix}
\begin{bmatrix}
\FF_1(t) \\
\FF_2(t) \\
\vdots \\
\FF_N(t)
\end{bmatrix}
\SSS(t) \\
+ \NN(t).   
\end{multline*}
Simplifying further, we have
\begin{equation}
\YY(t) = \HH(t) \FF(t) \SSS(t) + \NN(t),
\label{eq:2}
\end{equation}
where
\[
\HH(t) = \begin{bmatrix}
\HH_1(t) & \HH_2(t) & \cdots & \HH_N(t)
\end{bmatrix} ,
\]
and
\[
\FF(t) = \begin{bmatrix}
\FF_1(t)^\text{T} & \FF_2(t)^\text{T} & \cdots & \FF_N(t)^\text{T}
\end{bmatrix}^\text{T} .
\]
The \(N_t, N_r\) are considered as the number of transmit and receive antennas of the transceivers, respectively. The dimensions of \(\HH(t), \FF(t)\), and \(\SSS(t)\) are
$N_r \times N \cdot N_t$, $ N \cdot N_t \times N_s$, and $ N_s \times 1$, respectively,
where \(N_s\) is the maximum number of independent data streams that can be transmitted, given by $N_s = \min(N \cdot N_t, N_r)$.

\subsection{Ideal Scenario and Challenges}

In an ideal scenario, all transmitters are assumed to transmit at a specified time with identical carrier frequencies (synchronized oscillators). The precoder, \(\FF(t)\), is designed to maximize the received signal power at the receiver. However, in practical systems, imperfections in the oscillators at the transmitters result in misalignment of time references and carrier frequencies \cite{Zhou2008FrequencyA}. The precoder design typically involves transmitting pilot signals either from the receiver to all transmitters or from all transmitters to the receiver. These pilots are used to estimate the channel, based on which the precoder is computed. Nevertheless, this process introduces a delay in the dissemination of the precoder information to all transmitters. Additionally, due to the dynamic nature of wireless environments, the channel conditions at the time \(t\) of coherent joint transmission differ from those observed during the pilot transmission at time \((t - \tau)\), where \(\tau\) represents the feedback delay. These inconsistencies will significantly affect system performance, reducing received signal strength and causing inter-layer and inter-symbol interference (ISI).

\subsection{Impact of channel state information feedback delay}
Assuming perfect synchronization, the precoder is computed based on the channel at time \(t - \tau\). Consequently, the received signal at the receiver is expressed as
\[
\YY(t) = \HH(t) \FF(t - \tau) \SSS(t) + \NN(t),
\]
where \(\FF(t - \tau)\) is the delayed precoder. The elements of the channel \(\HH(t)\) exhibit the properties of a strict-sense stationary process, which satisfies the following
\[
\mathbb{E}[\HH_{i,j}(t)] = 0, \quad \mathbb{E}[\HH_{i,j}^\text{H}(t) \HH_{i,j}(t)] = 
\begin{cases}
1, & \text{if } i = j, \\
0, & \text{otherwise.}
\end{cases},
\]
\[
\rho_{i,j}(\tau) = \mathbb{E}[\HH_{i,j}^\text{H}(t) \HH_{i,j}(t - \tau)] =
\begin{cases}
J_0(2 \pi f_d \tau), & \text{if } i = j, \\
0, & \text{if } i \neq j.
\end{cases}
\]
where \(J_0\) is the zeroth-order Bessel function of the first kind, and \(f_d\) is the Doppler frequency shift. The channel \(\HH(t)\) can be decomposed using the auto-regressive process \cite{380120,1651496,AR_GRV} as 
\[
\HH(t) = \mathbf{E}(t) + \rho(\tau) \HH(t - \tau),
\]
where \(\EE(t)\) represents the residual complex Gaussian distributed matrix random process with its elements following 
\[\EE_{i,j}(t)\sim \mathcal{CN}(0, 1 - \rho^2(\tau))\]
and is independent of the channel $\HH(t - \tau)$. Substituting this into the signal model given in Eq.~\ref{eq:2}, we obtain
\[
\YY(t) = \EE(t) \FF(t - \tau) \SSS(t) + \rho(\tau) \HH(t - \tau) \FF(t - \tau) \SSS(t)  + \NN(t).    
\]
Here, the term \(\EE(t) \FF(t - \tau) \SSS(t)\) represents the interference component causing inter-layer interference since \(\EE(t)\) is unknown, and  the precoder built from \(\HH(t-\tau)\) is independent of $\EE(t)$, whereas \(\rho(\tau) \HH(t - \tau) \FF(t - \tau) \SSS(t)\) constitutes the signal component.

\subsection{Impact of synchronization errors}
The analysis assumes that synchronization among the transmitters has been performed, but residual time-frequency offsets are still present. The synchronization procedure involves one transmitter acting as the reference, to which all other transmitters synchronize in both time and frequency. Considering the impact of time and frequency synchronization errors, the received signal model is further modified to include independent time and frequency offsets
\[
\YY(t) = \sum_{i=1}^{N} \HH_i(t) \FF_i(t - \tau - T_i) \SSS(t - T_i) e^{j 2 \pi f_i (t - T_i)} + \NN(t),
\]
where \(T_i\) is the time offset and \(f_i\) is the frequency offset at the $i$-th transmitter. 
The frequency offset has two phase terms, one is the initial phase offset $- 2 \pi f_i T_i$, and the second is the phase rotation over time $ 2 \pi f_i t$. 
Assuming the phase rotations caused by frequency offsets $f_i$ are minimal during the coherent joint transmission, we have 
\[
\YY(t) = \sum_{i=1}^{N} \HH_i(t) \FF_i(t - \tau ) \SSS(t - T_i) e^{j \phi_i} + \NN(t),
\]
where $\phi_i$ is the initial random phase offset associated with $i$-th transmitter frequency offset $f_i$. Sampling the received signal with an optimum sampling interval \(T_s\), we obtain the discretized version of the received signal as
\begin{align}
\YY(k)  = \sum_{i=1}^{N} \HH_i(k) \FF_i(\bar{k}) \SSS(k - \eta_i) e^{j 2 \pi f_i k T_s} + \NN(k),
\label{eq:3}
\end{align}
where \(\eta_i = \frac{T_i}{T_s}\) is the normalized time offset and $\bar{k} = k - T$, where $T = \frac{\tau}{T_s}$ indicates delay in samples. Using linear interpolation, the transmitted signal \(\SSS[k - \eta_i]\) can be expressed as
\begin{align*}
\SSS(k - \eta_i) &= (1 - \eta_i) \SSS(k) + \eta_i \SSS(\Tilde{k}), \\
\text{where } \Tilde{k} &=
\begin{cases} 
k + 1, & \text{if } \eta_i < 0, \\ 
k - 1, & \text{if } \eta_i > 0.
\end{cases}
\end{align*}
Here, \( \eta_i < 0 \) indicates that data transmission from the corresponding transmitter started early, while \( \eta_i > 0 \) indicates that the data transmission was delayed with respect to a transmitter that acts as a reference. 

\section{Capacity Analysis}
The spectral efficiency is expressed as
\[
R = \log_2 \det \left( \mathbf{I} + \text{SINR} \right),
\]
where \(\text{SINR}\) is the Signal-to-Interference-plus-Noise Ratio, where
\[
\text{SINR} = \frac{\mathbf{S}}{\mathbf{I}_\text{int} + \mathbf{N}}
\]
is measuring the quality of the received signal $\mathbf{S}$ (Signal power) in the presence of both interference power $\mathbf{I}_\text{int}$ and noise variance $\mathbf{N}$. The interference arises from the non-coherent combining of the transmitted signals from multiple transmitters due to the synchronization offsets and CSI feedback delay. 



Expanding \(\SSS(k - \eta_i)\) in Eq.~\ref{eq:3}, we have the received signal 
\begin{multline*}
= \sum_{i=1}^N \HH_i(k) \FF_i(\bar{k}) \left[ (1 - \eta_i) \SSS(k)  
+ \eta_i \SSS(\bar{k}) \right] e^{j \phi_i}  + \NN(k)
\end{multline*}
\begin{multline*}
=  \sum_{i=1}^N \underbrace{ \HH_i(k) \FF_i(\bar{k}) (1 - \eta_i)  \SSS(k) e^{j \phi_i}}_{\text{Signal component with Inter-layer Interference}} + \\
 \sum_{i=1}^N  \underbrace{\HH_i(k) \FF_i(\bar{k}) \eta_i \SSS(\Tilde{k}) e^{j \phi_i}}_{\text{ISI component}} + \NN(k)    
\end{multline*}
Focusing on the signal component, moving the phase \(e^{j \phi_i}\) and time offsets \(\eta_i\) into the summation, and extracting \(\SSS(k)\) as a common factor, we have the Signal part as
\begin{multline*}
 \sum_{i=1}^N \HH_i(k) \FF_i(\bar{k}) (1 - \eta_i) \SSS(k) e^{j \phi_i} = \\
 \sum_{i=1}^N \HH_i(k) \mathbf{\theta}_i(k) \mathbf{\phi}_i(k) \FF_i(\bar{k}) \SSS(k), 
\end{multline*}
where \(\phi_i(k)\) and \(\theta_i(k)\) represents the diagonal matrices of size $N_T$ with phase offsets and time offsets respectively for the $i$-th transmitter. The phase and time offsets among the antennas of the $i$-th transmitter are equal because their RF chains are driven by a common oscillator. Finally, the received signal model with interference can be expressed as 
\begin{multline*}
\YY(k) = \underbrace{{\HH}(k) \theta(k) \phi(k) \FF(\bar{k}) \SSS(k) }_{\text{Signal component with Inter-layer Interference}} + \\
\underbrace{{\HH}(k) \psi(k) \phi(k) \FF(\bar{k}) \SSS(\Tilde{k})}_{\text{ISI component}} + \NN(k).    
\end{multline*}
The phase term \(\phi(k)\) is a diagonal matrix containing the phase offsets from all transmitter antennas, where each transmitter has identical phase offsets. The time offset term \(\theta(k)\) is a diagonal matrix containing \(1 - \eta_i\), while the other time offset term \(\psi(k)\) contains the corresponding time offsets \(\eta_i\). Because of the \(\phi(k)\), \(\theta(k)\), and delay in $\bar{k}$, the precoder cannot effectively align with the channel in the Signal component. As a result, the signals from independent data streams bleed into one another, causing inter-layer interference. We consider the SVD-based precoding where the $\FF(\bar{k})$ precoding matrix is designed using the first $N_s$ right singular vectors corresponding to the positive singular values. Because of the time offset term \(\psi(k)\), the symbols either from time index $k-1$ or $k+1$ interfere with current symbols, causing inter-symbol interference.

Considering only frequency synchronization errors, the received signal can be expressed as
\[
\YY(k) = \HH(k) \phi(k) \FF({k}) \SSS(k) + \NN(k),
\]
Applying receive beamforming \(\mathbf{U}^\text{H}(k)\), where
\[
\HH(k) = \mathbf{U}(k) \mathbf{\Sigma}(k) \mathbf{V}^\text{H}(k),
\]
we obtain
\[
\hat{y}(k) = \underbrace{\mathbf{\Sigma}(k) \mathbf{V}^\text{H}(k) \phi(k) \FF({k})}_{\Phi(k) } \SSS(k) + \hat{\NN}(k),
\]
where \(\hat{\NN}(k) = \mathbf{U}^\text{H}(k)\NN(k)\) follows the same distribution as \(\NN(k)\), since \(\mathbf{U}(k)\) is a unitary matrix. If there are no phase shifts, i.e., \(\Phi(k)\) is an identity matrix, then there is no inter-layer interference. However, due to \(\phi(k)\), the product \(\mathbf{U}^\text{H}(k) \phi(k) \FF(k) \neq \mathbf{I}\) results in inter-layer interference. The interference from other streams can be calculated using the matrix
\[
\Phi(k) = \mathbf{\Sigma}(k) \mathbf{V}^\text{H}(k) \phi(k) \mathbf{V}(k).
\]
In each row of the matrix, the diagonal element of \(\Phi(k)\) represents the received signal stream symbol, while the off-diagonal elements indicate the inter-layer interference caused by other stream symbols. The spectral efficiency in the presence of frequency offsets is given by
\[
\sum_{s=1}^{N_s} \log_2 \left( 1 + \frac{\lvert \Phi_{s,s}(k) \rvert^2}{\sum_{s' \neq s}^{N_s} \lvert \Phi_{s,s'}(k) \rvert^2 + \sigma^2} \right),
\]
where \(\Phi_{s,s}(k)\) represents the diagonal elements (signal components), \(\Phi_{s,s'}(k)\) represents the off-diagonal elements (interference components), and \(\sigma^2\) is the noise variance. 

Similarly, the capacity degradation in the presence of both time and frequency offsets with feedback delay can be derived. The spectral efficiency can be expressed as 
\begin{equation*}
    \sum_{s=1}^{N_s} \log_2 \left( 1 + \frac{\mathcal{A}}{\mathcal{B} + \mathcal{C} + \sigma^2} \right)
\end{equation*}
where $\mathcal{A} = \lvert \Phi_{s,s}(k) \rvert^2$ is the signal power in $s$-th stream, $\mathcal{B} = \sum_{s' \neq s} \lvert \Phi_{s,s'}(k) \rvert^2$ is inter-layer interference from the other streams due to frequency, time, and feedback delay offsets, and $\mathcal{C}= \sum_{s'}^{N_s} \lvert \zeta_{s,s'}(k) \rvert^2$ is the inter-symbol interference from time offsets.




\section{Simulation \& Results}
In this section, we present the performance plots based on the expressions derived earlier. The number of transmit and receive antennas at the transmitters and receiver is set to 2, with the number of transmitters \(N\) set to 5. The simulations were performed using MATLAB. The analysis considers that each transmitter's time and frequency offsets differ from one another. We also consider that the transmission power of each transmitter is equal and normalized so that the sum of individual powers is 0 dB. The channel elements between each transmitter and the receiver follow independent Rayleigh fading with unit variance.   

\begin{table}[h]
\caption{Capacity drop for different initial phase offset $\phi$ values in degrees.}
\centering
\begin{tabular}{|c|c|c|}
\hline
$\phi$ & SNR = 15 dB & SNR = 25 dB \\ \hline
$3.6^\circ$ & 0.91\% & 4.19\%\\ \hline
$18^\circ$ & 12.0\% & 24.97\%\\ \hline
$36^\circ$ & 26.65\% & 42.54\%\\ \hline
$45^\circ$ & 33.89\% & 48.38\%\\ \hline
$90^\circ$ & 62.51\% & 68.93\%\\ \hline
\end{tabular}
\label{table:phi_capacity}
\end{table}
\vspace{-2em}
\begin{figure}[ht]
  \centering
    \includegraphics[width=0.9\linewidth]{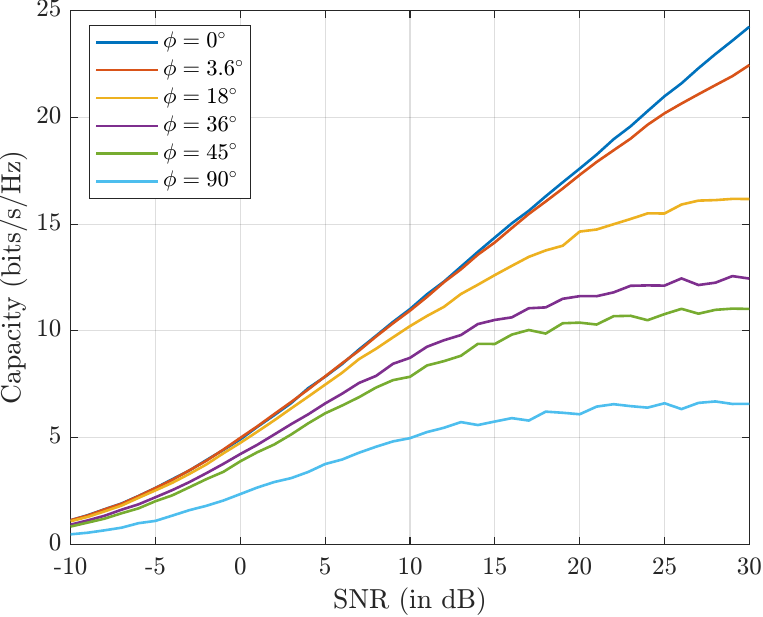}
    \caption{Capacity versus SNR (in dB) with different initial phase offsets $\phi$ arising due to residual frequency offsets}
  \label{fig:capacity_analysis_frequency_sync}
\end{figure}

Fig. \ref{fig:capacity_analysis_frequency_sync} shows the impact of different initial phase offsets on the capacity arising due to frequency offsets $f_i$. The phase offsets for each transmitter are different; however, they are derived from the Gaussian distribution with the same standard deviation $\phi$. The plot shows that as the initial phase offsets increase (arising due to frequency offsets), the capacity decreases. The capacity drops in percentages for SNR = 15 dB and SNR = 25 dB are tabulated in Table~\ref{table:phi_capacity}.

\begin{figure}[ht]
  \centering
  \begin{subfigure}{\columnwidth}
  \centering
    \includegraphics[width=0.9\linewidth]{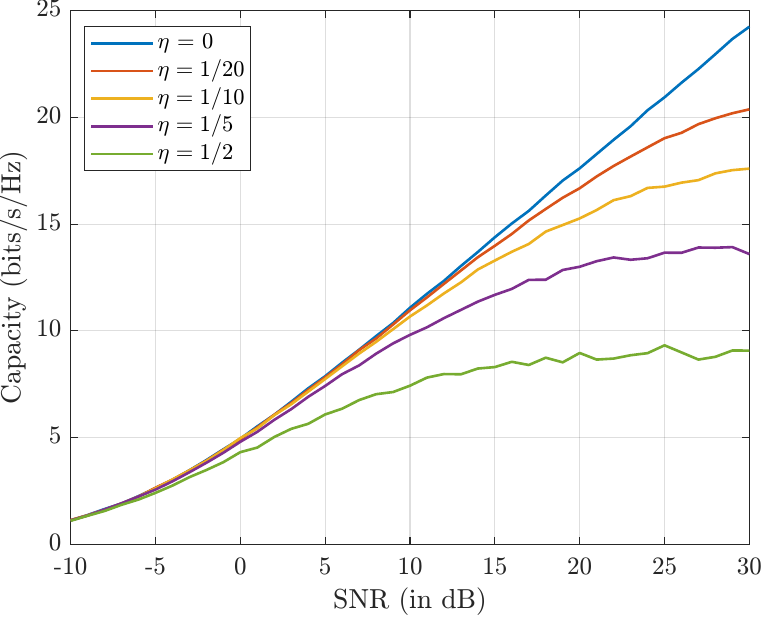}
    \caption{Capacity versus SNR (in dB) with different time offsets $\eta$}
    \label{fig:timesync_SNR_plot}
  \end{subfigure}
  \begin{subfigure}{\columnwidth}
  \centering
    \includegraphics[width=0.9\linewidth]{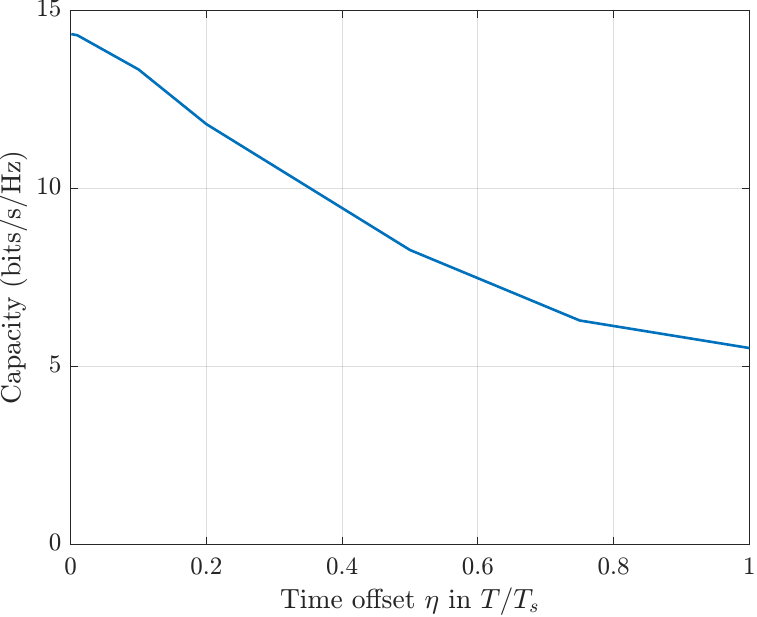}
    \caption{Capacity versus time offsets at SNR = 15 dB  }
    \label{fig:timesync_range_plot}
  \end{subfigure}
    \caption{Capacity analysis in the presence of time offsets $\eta$ with its distribution following Gaussian distribution}
  \label{fig:capacity_analysis_time_sync}  
\end{figure}

Fig. \ref{fig:capacity_analysis_time_sync} illustrates the system capacity in the presence of time offsets. The time offsets \(\eta\) are modeled as the Gaussian distribution with various standard deviation values ranging from 0 to \(1/2\). Fig. \ref{fig:timesync_SNR_plot} presents the capacity versus SNR plot, showing that at low SNRs, the impact of time offsets is minimal. Fig. \ref{fig:timesync_range_plot} depicts the capacity versus time offsets for values ranging from \(\eta = 0\) to one sample period (\(\eta = 1\)). As the time offset increases, the capacity decreases due to the loss of time synchronization. The capacity drops in percentages for SNR = 15 dB and SNR = 25 dB are tabulated in Table~\ref{table:eta_capacity}.

\begin{table}[h]
\centering
\caption{Capacity drop for different time offsets $\eta$ values.}
\begin{tabular}{|c|c|c|}
\hline
$\eta$ & SNR = 15 dB & SNR = 25 dB\\ \hline
$1/20$ & 2\% & 9.2\% \\ \hline
$1/10$ & 6.9\% & 19.95\% \\ \hline
$1/5$ & 18.21\% & 34.93\% \\ \hline
$1/2$ & 42.20\% & 56.66\% \\ \hline
\end{tabular}
\label{table:eta_capacity}
\end{table}
\vspace{-1em}

\begin{figure}[ht]
  \centering
  \begin{subfigure}{\columnwidth}
  \centering
    \includegraphics[width=0.9\linewidth]{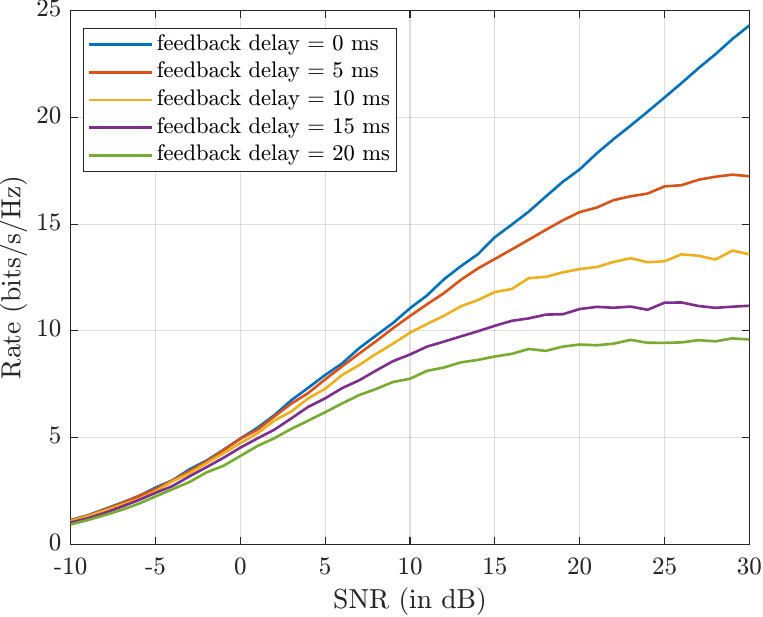}
    \caption{Capacity versus SNR (in dB) for $f_d$ = 5 Hz }
    \label{fig:feedback_delaySNR_plot}
  \end{subfigure}
  \begin{subfigure}{\columnwidth}
  \centering
    \includegraphics[width=0.9\linewidth]{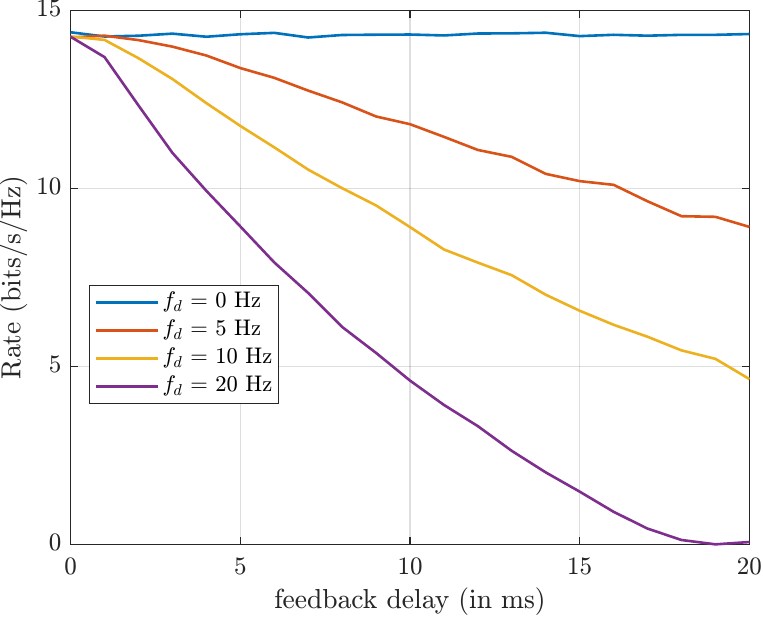}
    \caption{Capacity versus channel feedback delay (in ms) with different $f_d$}
    \label{fig:feedback_delay_fd_5_plot}
  \end{subfigure}

    \caption{Capacity analysis in the presence of the CSI Feedback Delay}

  \label{fig:capacity_analysis_feedback_delay}
\end{figure}

Fig. \ref{fig:capacity_analysis_feedback_delay} illustrates the system capacity in the presence of different feedback delays and Doppler spreads. Fig. \ref{fig:feedback_delaySNR_plot} presents the capacity versus SNR plot, highlighting the impact of a feedback delay for a range of values from 0 ms to 20 ms at a Doppler spread of 5 Hz. Fig. \ref{fig:feedback_delay_fd_5_plot} depicts capacity versus feedback delay for multiple Doppler frequencies. As the delay increases, the channel becomes increasingly de-correlated, leading to a significant reduction in capacity. Furthermore, it can be observed that the capacity degradation is more pronounced at higher Doppler frequencies. The capacity drops in percentages for SNR = 15 dB and SNR = 25 dB are tabulated in Table~\ref{table:tau_capacity}.

The tables and plots show that the capacity degradation due to offsets is significant under high SNR conditions. The noise impact is minimal, as it is low compared to the signal power. At high SNR, the plots show capacity performance flooring, mainly due to offsets.

\begin{table}[h]
\centering
\caption{Capacity drop for different feedback delays for $f_d$ = 5 Hz.}
\begin{tabular}{|c|c|c|}
\hline
$\tau$ in ms & SNR = 15 dB & SNR = 25 dB\\ \hline
$5$ & 4.08\% & 12.16\% \\ \hline
$10$ & 10.61\% & 25.42\% \\ \hline
$15$ & 19.08\% & 34.95\% \\ \hline
$20$ & 26.28\% & 42.76\% \\ \hline
\end{tabular}
\label{table:tau_capacity}
\end{table}
\vspace{-1.5em}
\vspace{-1em}

\section{Conclusion}
The impact of CSI feedback delay and time-frequency offsets in coherent joint transmission for distributed MIMO systems is investigated. The study focuses on transceivers with multiple antennas utilizing SVD-based precoding. The proposed analysis applies to any architecture performing coherent joint transmission. This study considered the transmit data and precoder information present at all the transmitters, not reflecting practical situations. The analysis shows that time and frequency offsets, along with CSI feedback delay, cause inter-layer interference. Additionally, time offsets result in inter-symbol interference. The amount of resources to ensure the transmit data and precoder information is present at all the transmitters should be accounted for in the capacity analysis, which reduces the overall capacity. 

Future work could explore the impact of synchronization errors in non-coherent joint transmission systems and investigate the effects of synchronization on cooperating receivers. Another area of interest involves determining the number of resources required to achieve the desired level of synchronization accuracy, as it eats up the resources for coherent joint transmission, and studying the trade-off when to switch to non-coherent joint transmission if the synchronization resources needed are high for coherent joint transmission.   

\section{Acknowledgement}
Efforts sponsored by the U.S. Government under the Training and Readiness Accelerator II (TReX II), OTA. The U.S. Government is authorized to reproduce and distribute reprints for Governmental purposes notwithstanding any copyright notation thereon. 

The views and conclusions contained herein are those of the authors and should not be interpreted as necessarily representing the official policies or endorsements, either expressed or implied, of the U.S. Government.

\bibliographystyle{IEEEtran}
\bibliography{reference}

\begin{thebibliography}{10}
\providecommand{\url}[1]{#1}
\csname url@samestyle\endcsname
\providecommand{\newblock}{\relax}
\providecommand{\bibinfo}[2]{#2}
\providecommand{\BIBentrySTDinterwordspacing}{\spaceskip=0pt\relax}
\providecommand{\BIBentryALTinterwordstretchfactor}{4}
\providecommand{\BIBentryALTinterwordspacing}{\spaceskip=\fontdimen2\font plus
\BIBentryALTinterwordstretchfactor\fontdimen3\font minus \fontdimen4\font\relax}
\providecommand{\BIBforeignlanguage}[2]{{%
\expandafter\ifx\csname l@#1\endcsname\relax
\typeout{** WARNING: IEEEtran.bst: No hyphenation pattern has been}%
\typeout{** loaded for the language `#1'. Using the pattern for}%
\typeout{** the default language instead.}%
\else
\language=\csname l@#1\endcsname
\fi
#2}}
\providecommand{\BIBdecl}{\relax}
\BIBdecl

\bibitem{ITU2024}
{International Telecommunication Union (ITU)}, ``{IMT}-2030 ({6G}) vision and research,'' \url{https://www.itu.int/en/ITU-R/study-groups/rsg5/rwp5d/imt-2030/Pages/default.aspx}, 2024, accessed: December 20, 2024.

\bibitem{You2021}
X.~You, C.-X. Wang, J.~Huang, and et~al., ``Towards {6G} wireless communication networks: vision, enabling technologies, and new paradigm shifts,'' \emph{Sci. China Inf. Sciences}, vol.~64, no. 110301, pp. 1--22, 2021.

\bibitem{9040264}
M.~Giordani, M.~Polese, M.~Mezzavilla, S.~Rangan, and M.~Zorzi, ``Toward {6G} networks: Use cases and technologies,'' \emph{IEEE Commun. Mag.}, vol.~58, no.~3, pp. 55--61, 2020.

\bibitem{10188355}
O.~Haliloglu, H.~Yu, C.~Madapatha, H.~Guo, F.~E. Kadan, A.~Wolfgang, R.~Puerta, P.~Frenger, and T.~Svensson, ``Distributed {MIMO} systems for {6G},'' in \emph{Joint European Conf. on Networks and Commun. \& 6G Summit (EuCNC/6G Summit)}, 2023, pp. 156--161.

\bibitem{10757642}
K.~S. Bondada, D.~J. Jakubisin, K.~Said, R.~M. Buehrer, and L.~Liu, ``Wireless mobile distributed-{MIMO} for {6G},'' in \emph{Proc. IEEE 100th Veh. Technol. Conf.}, 2024, pp. 1--7.

\bibitem{6601776}
X.~Hong, Y.~Jie, C.-X. Wang, J.~Shi, and X.~Ge, ``Energy-spectral efficiency trade-off in virtual {MIMO} cellular systems,'' \emph{IEEE J. Sel. Areas in Commun.}, vol.~31, no.~10, pp. 2128--2140, 2013.

\bibitem{6495775}
R.~Heath, S.~Peters, Y.~Wang, and J.~Zhang, ``A current perspective on distributed antenna systems for the downlink of cellular systems,'' \emph{IEEE Commun. Mag.}, vol.~51, no.~4, pp. 161--167, 2013.

\bibitem{6574665}
D.~N. Nguyen and M.~Krunz, ``Cooperative {MIMO} in wireless networks: recent developments and challenges,'' \emph{IEEE Net.}, vol.~27, no.~4, pp. 48--54, 2013.

\bibitem{6804225}
U.~Madhow, D.~R. Brown, S.~Dasgupta, and R.~Mudumbai, ``Distributed massive {MIMO}: Algorithms, architectures and concept systems,'' in \emph{Proc. Inf. Theory and Applications Workshop (ITA)}, 2014, pp. 1--7.

\bibitem{5490981}
D.~Castanheira and A.~Gameiro, ``Distributed antenna system capacity scaling [coordinated and distributed {MIMO}],'' \emph{IEEE Wireless Commun.}, vol.~17, no.~3, pp. 68--75, 2010.

\bibitem{1678166}
M.~Karakayali, G.~Foschini, and R.~Valenzuela, ``Network coordination for spectrally efficient communications in cellular systems,'' \emph{IEEE Wireless Commun.}, vol.~13, no.~4, pp. 56--61, 2006.

\bibitem{4114255}
W.~Choi and J.~G. Andrews, ``Downlink performance and capacity of distributed antenna systems in a multicell environment,'' \emph{IEEE Trans. Wireless Commun.}, vol.~6, no.~1, pp. 69--73, 2007.

\bibitem{5594708}
D.~Gesbert, S.~Hanly, H.~Huang, S.~Shamai~Shitz, O.~Simeone, and W.~Yu, ``Multi-cell {MIMO} cooperative networks: A new look at interference,'' \emph{IEEE J. Sel. Areas in Commun.}, vol.~28, no.~9, pp. 1380--1408, 2010.

\bibitem{5706317}
R.~Irmer, H.~Droste, P.~Marsch, M.~Grieger, G.~Fettweis, S.~Brueck, H.-P. Mayer, L.~Thiele, and V.~Jungnickel, ``Coordinated multipoint: Concepts, performance, and field trial results,'' \emph{IEEE Commun. Mag.}, vol.~49, no.~2, pp. 102--111, 2011.

\bibitem{7227028}
H.~Q. Ngo, A.~Ashikhmin, H.~Yang, E.~G. Larsson, and T.~L. Marzetta, ``Cell-free massive {MIMO}: Uniformly great service for everyone,'' in \emph{Proc. IEEE 16th Int. Workshop on Signal Processing Advances in Wireless Commun. (SPAWC)}, 2015, pp. 201--205.

\bibitem{7827017}
------, ``Cell-free massive {MIMO} versus small cells,'' \emph{IEEE Trans. Wireless Commun.}, vol.~16, no.~3, pp. 1834--1850, 2017.

\bibitem{10278803}
A.~Kumar, K.~Appaiah, and S.~R.~B. Pillai, ``On the ergodic sum capacity of multi-user mimo with distributed transmitter,'' in \emph{proc. IEEE Int. Conf. on Commun.}, 2023, pp. 2449--2454.

\bibitem{6735697}
D.~Scherber, P.~Bidigare, R.~ODonnell, M.~Rebholz, M.~Oyarzun, C.~Obranovich, W.~Kulp, D.~Chang, and D.~R.~B. III, ``Coherent distributed techniques for tactical radio networks: Enabling long range communications with reduced size, weight, power and cost,'' in \emph{Proc. IEEE MILCOM Conf.}, 2013, pp. 655--660.

\bibitem{6415392}
A.~Ozgur, O.~Leveque, and D.~Tse, ``Spatial degrees of freedom of large distributed {MIMO} systems and wireless ad hoc networks,'' \emph{IEEE J. Sel. Areas in Commun.}, vol.~31, no.~2, pp. 202--214, 2013.

\bibitem{9786576}
U.~Demirhan and A.~Alkhateeb, ``Enabling cell-free massive {MIMO} systems with wireless millimeter wave fronthaul,'' \emph{IEEE Trans. Wireless Commun.}, vol.~21, no.~11, pp. 9482--9496, 2022.

\bibitem{9048836}
S.~Schwarz and S.~Pratschner, ``Dynamic distributed antenna systems with wireless mmwave fronthaul,'' in \emph{Proc. 53rd Asilomar Conf. on Signals, Syst., and Comput.}, 2019, pp. 569--575.

\bibitem{10066319}
S.~Elhoushy, M.~Ibrahim, and W.~Hamouda, ``Downlink performance of cf massive {MIMO} under wireless-based fronthaul network,'' \emph{IEEE Trans. on Commun.}, vol.~71, no.~5, pp. 2632--2653, 2023.

\bibitem{9492307}
J.~A. Nanzer, S.~R. Mghabghab, S.~M. Ellison, and A.~Schlegel, ``Distributed phased arrays: Challenges and recent advances,'' \emph{IEEE Trans. Microw. Theory Tech.}, vol.~69, no.~11, pp. 4893--4907, 2021.

\bibitem{1417557}
S.~Jagannathan, H.~Aghajan, and A.~Goldsmith, ``The effect of time synchronization errors on the performance of cooperative miso systems,'' in \emph{Proc. IEEE GlobeCom Workshops.}, 2004, pp. 102--107.

\bibitem{4533899}
T.-D. Nguyen, O.~Berder, and O.~Sentieys, ``Impact of transmission synchronization error and cooperative reception techniques on the performance of cooperative {MIMO} systems,'' in \emph{2008 IEEE Int. Conf. on Commun.}, 2008, pp. 4601--4605.

\bibitem{6250567}
B.~Zafar, S.~Gherekhloo, F.~Roemer, and M.~Haardt, ``Impact of synchronization errors on alamouti-stbc-based cooperative {MIMO} schemes,'' in \emph{Proc. IEEE 7th Sensor Array and Multichannel Signal Processing Workshop (SAM)}, 2012, pp. 81--84.

\bibitem{4202181}
R.~Mudumbai, G.~Barriac, and U.~Madhow, ``On the feasibility of distributed beamforming in wireless networks,'' \emph{IEEE Trans. Wireless Commun.}, vol.~6, no.~5, pp. 1754--1763, 2007.

\bibitem{Zhou2008FrequencyA}
\BIBentryALTinterwordspacing
H.~Zhou, C.~Nicholls, T.~Kunz, and H.~M. Schwartz, ``Frequency accuracy \& stability dependencies of crystal oscillators,'' 2008. [Online]. Available: \url{https://api.semanticscholar.org/CorpusID:58408635}
\BIBentrySTDinterwordspacing

\bibitem{380120}
M.~Stojanovic, J.~Proakis, and J.~Catipovic, ``Analysis of the impact of channel estimation errors on the performance of a decision-feedback equalizer in fading multipath channels,'' \emph{IEEE Trans. on Commun.}, vol.~43, no. 2/3/4, pp. 877--886, 1995.

\bibitem{1651496}
Y.~Zhao, M.~Zhao, L.~Xiao, and J.~Wang, ``Capacity of time-varying rayleigh fading {MIMO} channels,'' in \emph{Proc. IEEE 16th Int. Symposium on Pers., Indoor \& Mobile Radio Commun.}, vol.~1, 2005, pp. 547--551.

\bibitem{AR_GRV}
{Herb Susmann}, ``Autoregressive processes are {G}aussian processes,'' \url{https://herbsusmann.com/2019/08/09/autoregressive-processes-are-gaussian-processes/}, 2024, accessed: December 20, 2024.

\end{thebibliography}


@online{dropmann2024golden,
  title={The Golden Bands and the Future of {6G}},
  author={Dropmann, Ulrich},
  year={2024},
  month={March},
  url={https://www.nokia.com/about-us/newsroom/articles/the-golden-bands-and-the-future-of-6g-0/},
    note         = {Accessed: 2025-01-21}
}


@INPROCEEDINGS{7227028,
  author={Ngo, Hien Quoc and Ashikhmin, Alexei and Yang, Hong and Larsson, Erik G. and Marzetta, Thomas L.},
  booktitle={Proc. IEEE 16th Int. Workshop on Signal Processing Advances in Wireless Commun. (SPAWC)}, 
  title={Cell-Free Massive {MIMO}: Uniformly great service for everyone}, 
  year={2015},
  volume={},
  number={},
  pages={201-205},
  keywords={MIMO;Power control;Correlation;Fading;Channel estimation;Throughput;Antennas},
  doi={10.1109/SPAWC.2015.7227028}}
@online{3gpp_sidelink_sTR,
    author       = {3GPP},
    title        = {Summary of Rel-18 Work Items},
    year         = 2024,
    url          = {https://www.3gpp.org/ftp/Specs/archive/21_series/21.918},
    note         = {Accessed: 2024-10-21}
}

@ARTICLE{9392777,
  author={Harounabadi, Mehdi and Soleymani, Dariush Mohammad and Bhadauria, Shubhangi and Leyh, Martin and Roth-Mandutz, Elke},
  journal={IEEE Commun. Standards Mag.}, 
  title={{V2X} in {3GPP} Standardization: {NR} Sidelink in Release-16 and Beyond}, 
  year={2021},
  volume={5},
  number={1},
  pages={12-21},
  keywords={Unicast;5G mobile communication;Safety;3GPP;Reliability;Vehicle-to-everything;Intelligent vehicles;Entertainment industry;Traffic control},
  doi={10.1109/MCOMSTD.001.2000070}}


@INPROCEEDINGS{10757642,
  author={Bondada, Kumar Sai and Jakubisin, Daniel J. and Said, Karim and Buehrer, R. Michael and Liu, Lingjia},
  booktitle={Proc. IEEE 100th Veh. Technol. Conf.}, 
  title={Wireless Mobile Distributed-{MIMO} for {6G}}, 
  year={2024},
  volume={},
  number={},
  pages={1-7},
}

@ARTICLE{You2021,
  author={You, Xue and Wang, Cheng-Xiang and Huang, Jie and et al.},
  title={Towards {6G} wireless communication networks: vision, enabling technologies, and new paradigm shifts},
  journal={Sci. China Inf. Sciences},
  volume={64},
  number={110301},
  pages={1-22},
  year={2021},
  doi={10.1007/s11432-020-2955-6}
}

@MISC{ITU2024,
  author={{International Telecommunication Union (ITU)}},
  title={{IMT}-2030 ({6G}) Vision and Research},
  year={2024},
  howpublished={\url{https://www.itu.int/en/ITU-R/study-groups/rsg5/rwp5d/imt-2030/Pages/default.aspx}},
  note={Accessed: December 20, 2024}
}


@ARTICLE{9040264,
  author={Giordani, Marco and Polese, Michele and Mezzavilla, Marco and Rangan, Sundeep and Zorzi, Michele},
  journal={IEEE Commun. Mag.}, 
  title={Toward {6G} Networks: Use Cases and Technologies}, 
  year={2020},
  volume={58},
  number={3},
  pages={55-61},
  keywords={6G mobile communication;5G mobile communication;Reliability;Wireless networks;Internet of Things;Intelligent sensors},
  doi={10.1109/MCOM.001.1900411}}



@online{sharma2024beyond,
    author       = {Vijay Sharma},
    title        = {Beyond Cell Towers: How 5G Sidelink Unleashes New Possibilities},
    year         = 2024,
    url          = {https://www.linkedin.com/pulse/beyond-cell-towers-how-5g-sidelink-unleashes-vijay-sharma-ykpic/},
    note         = {Accessed: 2024-10-21}
}


@misc{liu2024surveyrecentadvancesoptimization,
      title={A Survey of Recent Advances in Optimization Methods for Wireless Communications}, 
      author={Ya-Feng Liu and Tsung-Hui Chang and Mingyi Hong and Zheyu Wu and Anthony Man-Cho So and Eduard A. Jorswieck and Wei Yu},
      year={2024},
      eprint={2401.12025},
      archivePrefix={arXiv},
      primaryClass={cs.IT},
      url={https://arxiv.org/abs/2401.12025}, 
}


@article{article_Bengtsson,
author = {Bengtsson, M. and Ottersten, Björn},
year = {2001},
month = {01},
pages = {},
title = {Optimal and suboptimal transmit beamforming},
journal = {Handbook of Antennas in Wireless Commun.}
}

@INPROCEEDINGS{6364357,
  author={Zhu, Hao and Prasad, Narayan and Rangarajan, Sampath},
  booktitle={Proc. IEEE Int. Conf. on Commun. (ICC)}, 
  title={Precoder design for physical layer multicasting}, 
  year={2012},
  volume={},
  number={},
  pages={2140-2144},
  keywords={Approximation algorithms;Approximation methods;Multicast communication;Algorithm design and analysis;Physical layer;Optimization;Robustness},
  doi={10.1109/ICC.2012.6364357}}


@book{Heath,
place={Cambridge},
title={Foundations of {MIMO} Communication},
publisher={Cambridge University Press},
author={Heath Jr., Robert W. and Lozano, Angel},
year={2018}}



@INPROCEEDINGS{10188355,
  author={Haliloglu, Omer and Yu, Han and Madapatha, Charitha and Guo, Hao and Kadan, Fehmi Emre and Wolfgang, Andreas and Puerta, Rafael and Frenger, Pål and Svensson, Tommy},
  booktitle={Joint European Conf. on Networks and Commun. \& 6G Summit (EuCNC/6G Summit)}, 
  title={Distributed {MIMO} Systems for {6G}}, 
  year={2023},
  volume={},
  number={},
  pages={156-161},
  keywords={6G mobile communication;Wireless communication;Terminology;Simulation;Channel estimation;Standardization;Repeaters;6G;Distributed MIMO;D-MIMO;cell-free massive MIMO;5G;network controlled repeaters;NCR;integrated access and backhaul;reconfigurable intelligent surfaces;RIS;Ll mobility;non-coherent transmission;channel estimation;IAB;analog fronthaul},
  doi={10.1109/EuCNC/6GSummit58263.2023.10188355}}




@INPROCEEDINGS{7421222,
  author={Nayebi, Elina and Ashikhmin, Alexei and Marzetta, Thomas L. and Yang, Hong},
  booktitle={Proc. 49th Asilomar Conf. on Signals, Syst. and Comput.}, 
  title={Cell-Free Massive {MIMO} systems}, 
  year={2015},
  volume={},
  number={},
  pages={695-699},
  keywords={MIMO;Channel estimation;Optimization;Array signal processing;Resource management;Transmitting antennas},
  doi={10.1109/ACSSC.2015.7421222}}



@INPROCEEDINGS{1040690,
  author={Wonil Roh and Paulraj, A.},
  booktitle={Proc. IEEE 56th Veh. Tech. Conf.}, 
  title={{MIMO} channel capacity for the distributed antenna}, 
  year={2002},
  volume={2},
  number={},
  pages={706-709 vol.2},
  keywords={MIMO;Channel capacity;Fading;Microscopy;Downlink;Mobile antennas;Rayleigh channels;Covariance matrix;Mobile communication;Shadow mapping},
  doi={10.1109/VETECF.2002.1040690}}



@INPROCEEDINGS{1040381,
  author={Dohler, M. and Dominguez, J. and Aghvami, H.},
  booktitle={Proc. IEEE 56th Veh. Tech. Conf.}, 
  title={Link capacity analysis for virtual antenna arrays}, 
  year={2002},
  volume={1},
  number={},
  pages={440-443 vol.1},
  keywords={Antenna arrays;MIMO;Ad hoc networks;Power system relaying;Fading;Space time codes;Mobile antennas;Receiving antennas;Mobile handsets;Communication systems},
  doi={10.1109/VETECF.2002.1040381}}



@ARTICLE{1678166,
  author={Karakayali, M.K. and Foschini, G.J. and Valenzuela, R.A.},
  journal={IEEE Wireless Commun.}, 
  title={Network coordination for spectrally efficient communications in cellular systems}, 
  year={2006},
  volume={13},
  number={4},
  pages={56-61},
  keywords={Cellular networks;Mobile communication;Downlink;Mobile antennas;Receiving antennas;Interference;Antenna arrays;Base stations;Signal to noise ratio;Land mobile radio cellular systems},
  doi={10.1109/MWC.2006.1678166}}


@ARTICLE{4114255,
  author={Choi, Wan and Andrews, Jeffrey G.},
  journal={IEEE Trans. Wireless Commun.}, 
  title={Downlink performance and capacity of distributed antenna systems in a multicell environment}, 
  year={2007},
  volume={6},
  number={1},
  pages={69-73},
  keywords={Downlink;Interference;Transmitting antennas;Wireless communication;Radio frequency;Base stations;Performance analysis;Multiaccess communication;Computer simulation;Art},
  doi={10.1109/TWC.2007.05207}}


@ARTICLE{5490981,
  author={Castanheira, Daniel and Gameiro, Atilio},
  journal={IEEE Wireless Commun.}, 
  title={Distributed antenna system capacity scaling [Coordinated and Distributed {MIMO}]}, 
  year={2010},
  volume={17},
  number={3},
  pages={68-75},
  keywords={MIMO;Antennas and propagation;Transmitting antennas;Scattering;Content addressable storage;Diversity methods;Mobile antennas;Interference;Antenna measurements;Receiving antennas},
  doi={10.1109/MWC.2010.5490981}}



@ARTICLE{5594708,
  author={Gesbert, David and Hanly, Stephen and Huang, Howard and Shamai Shitz, Shlomo and Simeone, Osvaldo and Yu, Wei},
  journal={IEEE J. Sel. Areas in Commun.}, 
  title={Multi-Cell {MIMO} Cooperative Networks: A New Look at Interference}, 
  year={2010},
  volume={28},
  number={9},
  pages={1380-1408},
  keywords={MIMO;Interference;Base stations;Downlink;Mobile communication;Decoding;Relays;Cooperation;MIMO;cellular networks;relays;interference;beamforming;coordination;multi-cell;distributed},
  doi={10.1109/JSAC.2010.101202}}


@ARTICLE{5706317,
  author={Irmer, Ralf and Droste, Heinz and Marsch, Patrick and Grieger, Michael and Fettweis, Gerhard and Brueck, Stefan and Mayer, Hans-Peter and Thiele, Lars and Jungnickel, Volker},
  journal={IEEE Commun. Mag.}, 
  title={Coordinated multipoint: Concepts, performance, and field trial results}, 
  year={2011},
  volume={49},
  number={2},
  pages={102-111},
  keywords={Interference;Performance evaluation;Base stations;Throughput;Mobile communication;Downlink;MIMO;Spectral analysis},
  doi={10.1109/MCOM.2011.5706317}}


@ARTICLE{6495775,
  author={Heath, Robert and Peters, Steven and Wang, Yi and Zhang, Jiayin},
  journal={IEEE Commun. Mag.}, 
  title={A current perspective on distributed antenna systems for the downlink of cellular systems}, 
  year={2013},
  volume={51},
  number={4},
  pages={161-167},
  keywords={MIMO;Antenna arrays;Base stations;Interference;Transmitting antennas;Downlink;Cellular networks;Distributed antenna systems},
  doi={10.1109/MCOM.2013.6495775}}


@INPROCEEDINGS{6735697,
  author={Scherber, Dzulkifli and Bidigare, Patrick and ODonnell, Richard and Rebholz, Matthew and Oyarzun, Miguel and Obranovich, Charles and Kulp, William and Chang, Daniel and III, D. Richard Brown},
  booktitle={Proc. IEEE MILCOM Conf.}, 
  title={Coherent Distributed Techniques for Tactical Radio Networks: Enabling Long Range Communications with Reduced Size, Weight, Power and Cost}, 
  year={2013},
  volume={},
  number={},
  pages={655-660},
  keywords={Receivers;Radio transmitters;Array signal processing;Channel estimation;Military communication;Arrays;Timing;Communication equipment;Radio Communication;Wireless networks;Distributed algorithms;Time measurement;MIMO},
  doi={10.1109/MILCOM.2013.117}}



@ARTICLE{6574665,
  author={Nguyen, Diep N. and Krunz, Marwan},
  journal={IEEE Net.}, 
  title={Cooperative {MIMO} in wireless networks: recent developments and challenges}, 
  year={2013},
  volume={27},
  number={4},
  pages={48-54},
  keywords={MIMO;Wireless sensor networks;Interference;Ad hoc networks;Throughput;Cooperative networks;Mobile computing},
  doi={10.1109/MNET.2013.6574665}}


@INPROCEEDINGS{6804225,
  author={Madhow, U. and Brown, D. R. and Dasgupta, S. and Mudumbai, R.},
  booktitle={Proc. Inf. Theory and Applications Workshop (ITA)}, 
  title={Distributed massive {MIMO}: Algorithms, architectures and concept systems}, 
  year={2014},
  volume={},
  number={},
  pages={1-7},
  keywords={MIMO;Array signal processing;Frequency measurement;Phase measurement;Radio transmitters;Receivers;Synchronization},
  doi={10.1109/ITA.2014.6804225}}


@ARTICLE{6601776,
  author={Hong, Xuemin and Jie, Yu and Wang, Cheng-Xiang and Shi, Jianghong and Ge, Xiaohu},
  journal={IEEE J. Sel. Areas in Commun.}, 
  title={Energy-Spectral Efficiency Trade-Off in Virtual {MIMO} Cellular Systems}, 
  year={2013},
  volume={31},
  number={10},
  pages={2128-2140},
  keywords={MIMO;Interference;Signal to noise ratio;Protocols;Measurement;Relays;Resource management;Energy efficiency;spectral efficiency;virtual MIMO;adaptive resource allocation},
  doi={10.1109/JSAC.2013.131013}}

@ARTICLE{6415392,
  author={Ozgur, Ayfer and Leveque, Olivier and Tse, David},
  journal={IEEE J. Sel. Areas in Commun.}, 
  title={Spatial Degrees of Freedom of Large Distributed {MIMO} Systems and Wireless Ad Hoc Networks}, 
  year={2013},
  volume={31},
  number={2},
  pages={202-214},
  keywords={MIMO;Antenna arrays;Ad hoc networks;Receiving antennas;Approximation methods;Wireless networks;Spatial degrees of freedom;large scale MIMO;multi-user MIMO;virtual MIMO;distributed MIMO;wireless ad hoc networks;hierarchical cooperation;linear capacity scaling},
  doi={10.1109/JSAC.2013.130209}}



@ARTICLE{7827017,
  author={Ngo, Hien Quoc and Ashikhmin, Alexei and Yang, Hong and Larsson, Erik G. and Marzetta, Thomas L.},
  journal={IEEE Trans. Wireless Commun.}, 
  title={Cell-Free Massive {MIMO} Versus Small Cells}, 
  year={2017},
  volume={16},
  number={3},
  pages={1834-1850},
  keywords={MIMO;Uplink;Downlink;Power control;Fading channels;Antennas;Wireless communication;Cell-Free Massive MIMO system;conjugate beamforming;massive MIMO;network MIMO;small cell},
  doi={10.1109/TWC.2017.2655515}}


@INPROCEEDINGS{10278803,
  author={Kumar, Anil and Appaiah, Kumar and Pillai, Sibi Raj B},
  booktitle={proc. IEEE Int. Conf. on Commun.}, 
  title={On the Ergodic Sum Capacity of Multi-User MIMO with Distributed Transmitter}, 
  year={2023},
  volume={},
  number={},
  pages={2449-2454},
  keywords={Fading channels;Transmitters;Precoding;Massive MIMO;Mean square error methods;Optimization;Signal to noise ratio;Cell Free Massive MIMO;Ergodic Sum Capacity;Access Points;Broadcast Channel;Multiple Access Channel},
  doi={10.1109/ICC45041.2023.10278803}}


@ARTICLE{9492307,
  author={Nanzer, Jeffrey A. and Mghabghab, Serge R. and Ellison, Sean M. and Schlegel, Anton},
  journal={IEEE Trans. Microw. Theory Tech.}, 
  title={Distributed Phased Arrays: Challenges and Recent Advances}, 
  year={2021},
  volume={69},
  number={11},
  pages={4893-4907},
  keywords={Phased arrays;Array signal processing;Wireless communication;Wireless sensor networks;Synchronization;Radar;Radar remote sensing;Distributed arrays;distributed beamforming;phased arrays;radar;remote sensing},
  doi={10.1109/TMTT.2021.3092401}}


@INPROCEEDINGS{1417557,
  author={Jagannathan, S. and Aghajan, H. and Goldsmith, A.},
  booktitle={Proc. IEEE GlobeCom Workshops.}, 
  title={The effect of time synchronization errors on the performance of cooperative MISO systems}, 
  year={2004},
  volume={},
  number={},
  pages={102-107},
  keywords={Clocks;Jitter;MIMO;Synchronization;Fading;Wireless sensor networks;Transmitting antennas;Intersymbol interference;Degradation;Bit error rate},
  doi={10.1109/GLOCOMW.2004.1417557}}

@INPROCEEDINGS{4533899,
  author={Nguyen, T.-D. and Berder, O. and Sentieys, O.},
  booktitle={2008 IEEE Int. Conf. on Commun.}, 
  title={Impact of Transmission Synchronization Error and Cooperative Reception Techniques on the Performance of Cooperative {MIMO} Systems}, 
  year={2008},
  volume={},
  number={},
  pages={4601-4605},
  keywords={MIMO;Wireless sensor networks;Degradation;Diversity methods;Energy consumption;Cooperative systems;Additive noise;Signal generators;Intersymbol interference;Channel state information},
  doi={10.1109/ICC.2008.863}}



@INPROCEEDINGS{6250567,
  author={Zafar, Bilal and Gherekhloo, Soheyl and Roemer, Florian and Haardt, Martin},
  booktitle={Proc. IEEE 7th Sensor Array and Multichannel Signal Processing Workshop (SAM)}, 
  title={Impact of synchronization errors on Alamouti-STBC-based cooperative {MIMO} schemes}, 
  year={2012},
  volume={},
  number={},
  pages={81-84},
  keywords={Signal to noise ratio;Interference;Synchronization;MIMO;Bit error rate;Vectors;Receivers},
  doi={10.1109/SAM.2012.6250567}}


@ARTICLE{4202181,
  author={Mudumbai, R. and Barriac, G. and Madhow, U.},
  journal={IEEE Trans. Wireless Commun.}, 
  title={On the Feasibility of Distributed Beamforming in Wireless Networks}, 
  year={2007},
  volume={6},
  number={5},
  pages={1754-1763},
  keywords={Array signal processing;Wireless networks;Frequency synchronization;Transmitters;Base stations;Wireless sensor networks;Master-slave;Energy efficiency;Transmitting antennas;Antenna arrays},
  doi={10.1109/TWC.2007.360377}}


@INPROCEEDINGS{10238024,
  author={Merlo, Jason M. and Nanzer, Jeffrey A.},
  booktitle={IEEE Int. Symposium on Antennas and Propagation and USNC-URSI Radio Sci. Meeting}, 
  title={Wireless Time and Phase Alignment for Wideband Beamforming in Distributed Phased Arrays}, 
  year={2023},
  volume={},
  number={},
  pages={365-366},
  keywords={Wireless communication;Phased arrays;Frequency synthesizers;Array signal processing;Conf.s;Synchronization;Broadband antennas},
  doi={10.1109/USNC-URSI52151.2023.10238024}}

@INPROCEEDINGS{7218555,
  author={Abari, Omid and Rahul, Hariharan and Katabi, Dina and Pant, Mondira},
  booktitle={Proc. IEEE Conf. Comput. Commun.}, 
  title={AirShare: Distributed coherent transmission made seamless}, 
  year={2015},
  volume={},
  number={},
  pages={1742-1750},
  keywords={Clocks;Wireless communication;Phase locked loops;Wireless sensor networks;Protocols;MIMO;Radio transmitters},
  doi={10.1109/INFOCOM.2015.7218555}}


@inproceedings{3448623,
author = {Alemdar, Kubra and Varshney, Divashree and Mohanti, Subhramoy and Muncuk, Ufuk and Chowdhury, Kaushik},
title = {RFClock: timing, phase and frequency synchronization for distributed wireless networks},
year = {2021},
isbn = {9781450383424},
publisher = {Association for Computing Machinery},
address = {New York, NY, USA},
url = {https://doi.org/10.1145/3447993.3448623},
doi = {10.1145/3447993.3448623},
booktitle = {Proc. 27th Annual Int. Conf. on Mobile Comput. and Netw.},
pages = {15–27},
numpages = {13},
location = {New Orleans, Louisiana},
series = {MobiCom '21}
}


@INPROCEEDINGS{8742232,
  author={Yan, Han and Hanna, Samer and Balke, Kevin and Gupta, Riten and Cabric, Danijela},
  booktitle={Proc. IEEE Aerosp. Conf.}, 
  title={Software Defined Radio Implementation of Carrier and Timing Synchronization for Distributed Arrays}, 
  year={2019},
  volume={},
  number={},
  pages={1-12},
  keywords={Synchronization;Array signal processing;Protocols;Frequency synchronization;OFDM;Radio transmitters},
  doi={10.1109/AERO.2019.8742232}}


@INPROCEEDINGS{6624252,
  author={Brown, D. Richard and Klein, Andrew G.},
  booktitle={Proc. 47th Annual Conf. on Inf. Sci. and Syst. (CISS)}, 
  title={Precise timestamp-free network synchronization}, 
  year={2013},
  volume={},
  number={},
  pages={1-6},
  keywords={Synchronization;Protocols;Clocks;Oscillators;Accuracy;Delay estimation;Wireless sensor networks;synchronization;delay estimation;oscillator dynamics;wireless communication;distributed communication systems},
  doi={10.1109/CISS.2013.6624252}}

@ARTICLE{9994246,
  author={Merlo, Jason M. and Mghabghab, Serge R. and Nanzer, Jeffrey A.},
  journal={IEEE Trans. Microw. Theory Tech.}, 
  title={Wireless Picosecond Time Synchronization for Distributed Antenna Arrays}, 
  year={2023},
  volume={71},
  number={4},
  pages={1720-1731},
  keywords={Synchronization;Wireless communication;Wireless sensor networks;Antenna arrays;Time-frequency analysis;Bandwidth;Clocks;Clock synchronization;distributed arrays;distributed beamforming;radar;remote sensing;two-way time transfer;wireless sensor networks;wireless synchronization},
  doi={10.1109/TMTT.2022.3227878}}


@INPROCEEDINGS{8396683,
  author={Kowalski, Daniel R. and Christman, Timothy M. and Klein, Andrew G. and Overdick, Mitchell W.S. and Canfield, Joseph E. and Richard Brown, D.},
  booktitle={Proc. IEEE Aerospace Conf.}, 
  title={Implementation and testing of a low-overhead network synchronization protocol}, 
  year={2018},
  volume={},
  number={},
  pages={1-8},
  keywords={Synchronization;Protocols;Clocks;RF signals;Software;Delay estimation},
  doi={10.1109/AERO.2018.8396683}}


@INPROCEEDINGS{9048836,
  author={Schwarz, Stefan and Pratschner, Stefan},
  booktitle={Proc. 53rd Asilomar Conf. on Signals, Syst., and Comput.}, 
  title={Dynamic Distributed Antenna Systems with Wireless mmWave Fronthaul}, 
  year={2019},
  volume={},
  number={},
  pages={569-575},
  keywords={Fading channels;Antenna arrays;Downlink;Wireless communication;Cellular networks;Quantization (signal);cloud RAN;distributed antenna systems;mmWave;optimization;cell-free MIMO},
  doi={10.1109/IEEECONF44664.2019.9048836}}


@ARTICLE{9786576,
  author={Demirhan, Umut and Alkhateeb, Ahmed},
  journal={IEEE Trans. Wireless Commun.}, 
  title={Enabling Cell-Free Massive {MIMO} Systems With Wireless Millimeter Wave Fronthaul}, 
  year={2022},
  volume={21},
  number={11},
  pages={9482-9496},
  keywords={Massive MIMO;Wireless communication;Computer architecture;Synchronization;Costs;Bandwidth;Central Processing Unit;Cell free massive MIMO;millimeter wave fronthaul;beamforming},
  doi={10.1109/TWC.2022.3177186}}

@ARTICLE{10066319,
  author={Elhoushy, Salah and Ibrahim, Mohamed and Hamouda, Walaa},
  journal={IEEE Trans. on Commun.}, 
  title={Downlink Performance of CF Massive {MIMO} Under Wireless-Based Fronthaul Network}, 
  year={2023},
  volume={71},
  number={5},
  pages={2632-2653},
  keywords={Wireless communication;Precoding;Millimeter wave communication;Microwave communication;Systems operation;Decoding;Array signal processing;Cell-free massive MIMO;centralized operation;distributed operation;downlink data rates;mmWave;wireless fronthaul network},
  doi={10.1109/TCOMM.2023.3255906}}

@ARTICLE{650240,
  author={Schmidl, T.M. and Cox, D.C.},
  journal={IEEE Trans. on Commun.}, 
  title={Robust frequency and timing synchronization for OFDM}, 
  year={1997},
  volume={45},
  number={12},
  pages={1613-1621},
  keywords={Robustness;Timing;Frequency synchronization;OFDM;Frequency estimation;Frequency division multiplexing;Wireless LAN;Sampling methods;Frequency domain analysis;Signal detection},
  doi={10.1109/26.650240}}


@INPROCEEDINGS{10017959,
  author={Polydoros, Andreas and Köse, Cenk},
  booktitle={Proc. IEEE MILCOM Conf.}, 
  title={Clustered Distributed Spatial Multiplexing}, 
  year={2022},
  volume={},
  number={},
  pages={938-943},
  keywords={Radio frequency;Military communication;Cooperative communication;Receivers;Ad hoc networks;Topology;Space division multiplexing;Distributed Spatial Multiplexing;Space-Division Multiple Access;MANET;virtual MIMO;Beamforming;SVD},
  doi={10.1109/MILCOM55135.2022.10017959}}


@ARTICLE{4785387,
  author={Mudumbai, Raghuraman and Brown Iii, D. Richard and Madhow, Upamanyu and Poor, H. Vincent},
  journal={IEEE Commun. Mag.}, 
  title={Distributed transmit beamforming: challenges and recent progress}, 
  year={2009},
  volume={47},
  number={2},
  pages={102-110},
  keywords={Array signal processing;Signal to noise ratio;Noise figure;Frequency synchronization;Prototypes;Antennas and propagation;Transmitting antennas;Frequency measurement;Wavelength measurement;Unmanned aerial vehicles},
  doi={10.1109/MCOM.2009.4785387}}



@INPROCEEDINGS{6189962,
  author={Mudumbai, Raghu and Madhow, Upamanyu and Brown, Rick and Bidigare, Patrick},
  booktitle={Proc. 44th Asilomar Conf. on Signals, Syst., and Comput.}, 
  title={{DSP}-centric algorithms for distributed transmit beamforming}, 
  year={2011},
  volume={},
  number={},
  pages={93-98},
  keywords={Array signal processing;Frequency estimation;Synchronization;Oscillators;Training;Baseband;Receivers;distributed beamforming;synchronization;baseband algorithms},
  doi={10.1109/ACSSC.2011.6189962}}

@INPROCEEDINGS{6319679,
  author={Brown, D. Richard and Bidigare, Patrick and Dasgupta, Soura and Madhow, Upamanyu},
  booktitle={Proc. IEEE Statistical Signal Processing Workshop (SSP)}, 
  title={Receiver-coordinated zero-forcing distributed transmit nullforming}, 
  year={2012},
  volume={},
  number={},
  pages={269-272},
  keywords={Receivers;Vectors;Noise;Array signal processing;Protocols;Channel estimation;Local oscillators;cooperative communication;distributed transmission;feedback systems;oscillator dynamics;tracking},
  doi={10.1109/SSP.2012.6319679}}

@ARTICLE{tr38785,
  author={3GPP},
  journal={TR 38.785}, 
  title={{User Equipment (UE) radio transmission and reception for enhanced NR sidelink}}, 
  year={2022}}



@inproceedings{Zhou2008FrequencyA,
  title={Frequency Accuracy \& Stability Dependencies of Crystal Oscillators},
  author={Hui Zhou and C.W.T. Nicholls and Thomas Kunz and Howard M. Schwartz},
  year={2008},
  url={https://api.semanticscholar.org/CorpusID:58408635}
}

@ARTICLE{380120,
  author={Stojanovic, M. and Proakis, J.G. and Catipovic, J.A.},
  journal={IEEE Trans. on Commun.}, 
  title={Analysis of the impact of channel estimation errors on the performance of a decision-feedback equalizer in fading multipath channels}, 
  year={1995},
  volume={43},
  number={2/3/4},
  pages={877-886},
  keywords={Channel estimation;Decision feedback equalizers;Fading;Adaptive algorithm;Performance analysis;Error probability;Estimation error;Covariance matrix;Error analysis},
  doi={10.1109/26.380120}}


@INPROCEEDINGS{1651496,
  author={Yifei Zhao and Ming Zhao and Limin Xiao and Jing Wang},
  booktitle={Proc. IEEE 16th Int. Symposium on Pers., Indoor \& Mobile Radio Commun.}, 
  title={Capacity of Time-Varying Rayleigh Fading {MIMO} Channels}, 
  year={2005},
  volume={1},
  number={},
  pages={547-551},
  keywords={Rayleigh channels;{MIMO};Fading;Time varying systems;Frequency;Interleaved codes;Gaussian channels;Power system modeling;Mean square error methods;Channel estimation;{MIMO} systems;Channel capacity;Time-varying channels;Channel estimation},
  doi={10.1109/PIMRC.2005.1651496}}

@MISC{AR_GRV,
  author={{Herb Susmann}},
  title={Autoregressive Processes are {G}aussian Processes},
  year={2024},
  howpublished={\url{https://herbsusmann.com/2019/08/09/autoregressive-processes-are-gaussian-processes/}},
  note={Accessed: December 20, 2024}
}

\end{document}